\pdfoutput=1

\documentclass[11pt]{article}

\usepackage{scrextend}

\usepackage{NLPAICS2026}
\usepackage{times}
\usepackage{latexsym}

\usepackage[T1]{fontenc}

\usepackage[utf8]{inputenc}

\usepackage{microtype}

\usepackage{inconsolata}

\usepackage[group-separator={,}]{siunitx}
\usepackage{adjustbox}
\usepackage{listings}
\usepackage{hyperref}
\usepackage{float}
\usepackage{censor}

\newcommand*\rot[1]{\rotatebox{25}{#1}}

\title{{T}ransformer-{A}ssisted {L}{L}{M}-{B}ased {S}ource {C}ode {S}ummarisation: to {E}nable {M}ore {S}ecure {S}oftware {D}evelopment}

\author{
Jesse Phillips,$^{1}$ Tracy Hall,$^{1}$ Paul Rayson,$^{1,2}$ \and Mo El-Haj$^{1,2}$\\[0.5em]
$^{1}$School of Computing and Communications, Lancaster University, UK\\
$^{2}$College of Engineering \& Computer Science, VinUniversity, Vietnam\\[0.5em]
\texttt{\{j.m.phillips, tracy.hall, p.rayson, m.el-haj\}@lancaster.ac.uk}\\
\texttt{\{paul.r2, elhaj.m\}@vinuni.edu.vn}
}

\begin{document}
\maketitle
\begin{abstract}
Neural Source Code Summarisation (NSCS) aims to generate natural language summaries of source code to improve developer and maintainer understanding of code.  Source code summaries are vital for the maintenance phase of the Secure Software Development Lifecycle (SSDLC) as they improve maintainers' understanding of code, in order to reduce the number of bugs and vulnerabilities in a software system.  However, summaries are often missing, incomplete, or outdated in many software systems.  Solutions to this problem use small, task-specific Transformer models or code-aware Large Language Models (LLMs). Task-specific Transformer-generated summaries often score well across many NLG metrics but these NLG metrics reward lexical overlap, rather than summary quality.  
Conversely, LLMs’ ability to capture semantics in order to produce high-quality summaries 
presents an exciting solution to this problem, especially with the increased availability of LLMs and the increase in capability of workstation hardware over recent years meaning that some LLMs can be run from developers' workstations.  However, LLM summaries of code often differ greatly from developer-written summaries in terms of the words and phrases used due to the abstractive nature of LLMs, resulting in low scores across NLG metrics.  We show how combining these two methods by using Transformer-generated summaries in prompt engineering may enable LLMs to create better source code summaries in order to better enable software practitioners to maintain secure systems.  We prompt four LLMs, using four different prompts - with the use of a task-specific Transformer to aid the LLMs in the prompts.  We present ``Transformer-Assisted LLM-Based Source Code Summarisation'' - a method through which, we observe an improvement of 7.8\% \textsc{Bleu}-4 and 5\% \textsc{Bert}Score on CodeLlama.
\end{abstract}

\section{Introduction}\label{sec:Introduction}
Source code summarisation is a key part of the software documentation process, which aims to make source code methods more understandable to software practitioners, especially during software maintenance.  Since 2020, the United Kingdom's National Cyber Security Centre have recommended software practitioners ``document and comment clearly and consistently''\footnote{\href{https://web.archive.org/web/20260315192904/https://www.ncsc.gov.uk/collection/developers-collection/principles/produce-clean-maintainable-code}{ncsc.gov.uk/collection/developers-collection/principles/produce-clean-maintainable-code}}
as part of secure software development.  \citet{khan-etal-2022-systematic} identify a lack of secure development or coding as part of an extensive list of security risks in practice.  However, in an increasingly fast-paced, Agile working environment, documentation is often a neglected task due to being seen as a low-priority under time constraints.  The Agile Manifesto is built on a series of values, one of which being: ``Working software over comprehensive documentation.'' \citep{fowler-etal-2001-agile}, which means that documentation tasks often get rushed, or ignored.  Despite this, developers find that documentation is of importance, with \citet{venigalla-chimalakonda-2021-understanding} finding that ``Software documentation aids better project comprehension[...]''.

The motivation behind this paper is to address the problems that insufficient source code summarisation can cause in secure software development.

Neural Source Code Summarisation (NSCS) aims to solve this problem by reducing the load on the developer and automating the task of writing code summaries.  Work in this field often relies on small-scale, task-specific Transformer models, as shown by \citet{ahmad-etal-2020-transformer} and \citet{feng-etal-2020-codebert}, or using LLMs, as shown by  \citet{ahmed-devanbu-2022-few} and \citet{geng-etal-2024-large}.  While these approaches are both built on Transformer architectures, the key differences between these two approaches are the size of the models and the task they are trained to perform, with task-specific Transformers trained solely on datasets of code and related summaries, and LLMs trained on entire corpora of language in order to build a model which can be instructed to perform a variety of tasks using Natural Language.

The aim of this research is to combine these two approaches in order to achieve a ``best of both worlds'' result, with the fluency of language produced by an LLM, and the use of technical key words, produced by a task-specific Transformer.  To this aim, we propose the following research questions:

\begin{addmargin}[16pt]{0pt}
    \begin{itemize}
        \item [\textbf{RQ. 1:}] Does prompting a Large Language Model with the output of a smaller Transformer model improve the summary quality when evaluated against NLG metrics?

        
        \item [\textbf{RQ. 2:}] As summarisation is a specific NLP task - which LLMs are pretrained to be able to perform - does referencing summarisation in the prompt affect the summary quality?

        
        \item [\textbf{RQ. 3:}] Does using a code-aware LLM provide an improvement over an English-language LLM when evaluated on an NSCS task?

    \end{itemize}
\end{addmargin}

\section{Related Work}
LLMs have provided new avenues of research in NSCS in recent years. In 2022, \citet{ahmed-devanbu-2022-few} proposed a few-shot approach to prompting LLMs for an NSCS task, using a fill-in-the-blank approach on Codex, finding improved performance when compared to Code\textsc{Bert}, GraphCode\textsc{Bert}, and CodeT5 - however, this is not a like-for-like comparison as Codex uses a few-shot prompt setting, where the other models are smaller-scale task-specific Transformer models.  Similarly, \citet{geng-etal-2024-large}, used Codex' Code-DaVinci-002 model for code comment generation.  The model is evaluated on both \citet{leclair-mcmillan-2019-recommendations}'s Funcom dataset and \citet{hu-etal-2018-summarizing}'s TL-CodeSum dataset.

\citet{ahmed-etal-2024-automatic} builds on \citet{ahmed-devanbu-2022-few}'s previous work with a new prompting methodology, ``Automatic Semantic Augmentation of Prompts (ASAP)'', which proposes the use of ``what developers think about'' as part of the prompt, by adding information such as the repository name and path, or tagged identifiers such as variable names and scopes.  The ASAP prompting method is used primarily on Code-DaVinci-002, and also tested on Text-DaVinci-003 and GPT-3.5-Turbo.


\citet{haldar-hockenmaier-2024-analyzing} compare Llama 2 and PaLM 2 LLMs to CodeT5 on a code summarisation task, using both \textsc{Bleu}-4 and \textsc{Bert}Score metrics on the CodeXGLUE dataset.  To verify the performance, the authors perform this on both the standard dataset, as well as augmented methods from the dataset with obfuscated function names, adversarial function names, missing code structure, and missing function bodies.  

\citet{sun-etal-2025-source} test CodeLlama, StarChat-$\beta$,  GPT-3.5, and GPT-4 using zero-shot, chain-of-thought, and expert prompting methods.  The authors use \textsc{Bleu}, \textsc{Meteor}, and \textsc{Rouge}-L metrics to compare these models, as well as building LLM-based metrics (similar to \textsc{Bert}Score) based on each model in order to evaluate them.  In this evaluation, the authors also find a correlation between human evaluation and LLM-based evaluation when analysing reference summaries, suggesting that LLM-based evaluation metrics are good indicators of summary quality.

In 2024, \citet{phillips-etal-2024-metric} proposed the CodeSum\textsc{Bart} model, a task-specific, small-scale Transformer NSCS model ``to enable more secure software development''.

In this paper, we build on \citeposs{phillips-etal-2024-metric} principle that generating source code summaries aids secure development practices, and on recent works in using LLMs for NSCS in order to develops a method of prompting LLMs to better perform NSCS tasks for secure development.


\section{Dataset}\label{sec:Dataset}
The primary dataset used for this research was \citeposs{leclair-mcmillan-2019-recommendations} Funcom Dataset, processed using an updated version of \citeposs{phillips-etal-2022-improved} dataset cleaning methodology.  To improve upon \citeposs{phillips-etal-2024-metric} cleaning methodology, we manually searched the dataset for tokenisation errors, in order to find any potential deficiencies in the methodology which could be improved upon.  We found the following repeated errors:
\begin{itemize}\setlength\itemsep{-4px}
    \item Phrases in snakecase being parsed as a single token,
    \item Punctuation attached to words in tokens, including apostrophes in use as single quotation marks
    \item Various spacing errors.
\end{itemize}

In order to rectify this, we added a new tokenisation method to the Java Class used to preprocess the data, which can be found in Appendix \ref{sec:AppendixJava}.
In order to measure the improvement this had on the dataset, we compared a list of all tokens in the dataset against a corpus of English Language.  In the initial dataset, we found \num{146905} out-of-dictionary tokens.  After performing these improvements to the tokenisation, We found \num{40941} out-of-dictionary tokens - a reduction of \num{105964}.  When manually checking these tokens, they appear to be largely technical jargon, with some project-specific terms.

\begin{table*}[!h]
    \centering
    \begin{tabular}{p{.16\linewidth}rrrrrr}
        \hline
        Dataset cleaning method used & \rot{\textsc{Bleu}-1} & \rot{\textsc{Bleu}-4} & \rot{Smoothed \textsc{Bleu}-4} & \rot{\textsc{Meteor}} & \rot{FrugalScore} & \rot{\textsc{Bert}Score} \\
        \hline
        Updated method & \textbf{55.79} & \textbf{33.57} & \textbf{33.79} & \textbf{26.99} & \textbf{74.19} & \textbf{72.10} \\
        Original method & 53.58 & 30.41 & 30.66 & 24.96 & 73.59 & 71.70 \\
    \end{tabular}
    \caption{Improvements to CodeSum\textsc{Bart} Caused by an Updated Dataset Cleaning Method}\label{table:UpdatedDataset}
\end{table*}

The Training and Validation splits from this dataset were used to train a CodeSum\textsc{Bart} \citep{phillips-etal-2024-metric}. The trained model is then used to generate example summaries for one-shot LLM prompts.

Table \ref{table:UpdatedDataset} shows the improvement in performance caused by training a CodeSum\textsc{Bart} model using this dataset cleaning method.  By using the updated cleaning method created by appending the Java method in Appendix \ref{sec:AppendixJava} to \citeposs{phillips-etal-2022-improved} Java Dataset Cleaner, we observe a small improvement in all evaluation metrics used, when compared to \citeposs{phillips-etal-2022-improved} dataset.

As shown in Table \ref{table:DatasetSplit}, the final dataset contained \num{499997} Java method - source code summary pairs, with each method and summary tokenised and cleaned in order to better support model training.
\begin{table}[!ht]
    \centering
    \begin{tabular}{rrr}
      \hline
      Training & Validation & Evaluation \\
      \hline
      \num{399999} & \num{49999} & \num{49999} \\
      80\% & 10\% & 10\% \\
      \hline
    \end{tabular}
    \caption{Split of methods in the dataset.}\label{table:DatasetSplit}
\end{table}

\section{Methodology}
In order to answer the research questions in Section \ref{sec:Introduction}, we generated a series of prompts, with one zero-shot prompt, and three one-shot prompts, with an example provided by a small-scale task-specific Transformer model.  We evaluate the performance of four open source LLMs, including two code-aware LLMs, using these four prompts, on the dataset described in Section \ref{sec:Dataset}.  In order to perform this evaluation, we use three variants of the \textsc{Bleu} metric, \textsc{Rouge}-L, and \textsc{Bert}Score.

\subsection{Methodology for RQ. 1}\label{sec:MethodologyRQ1}
We began by comparing a zero-shot prompt to a one-shot prompt using the outputs of a CodeSum\textsc{Bart} model on the same task to generate the examples for the one-shot prompting.  We used the Evaluation split from the dataset for this task.  Google's LLM Prompt Engineering guide\footnote{\href{https://www.kaggle.com/whitepaper-prompt-engineering}{kaggle.com/whitepaper-prompt-engineering}}
defines a zero-shot prompt as ``[...] the simplest type of prompt. It only provides a description of a task and some text for the LLM to get started with.'', and a one-shot prompt as a prompt which ``[...] provides a single example, hence the name one-shot. The idea is that the model has an example it can imitate to best complete the task.''.  These are the definitions we adopted for this work.  

\citet{giray-2023-prompt} break down a prompt into four parts: instruction, context, input data, and output indicator.  ``Instruction'' gives the specific task to tell the model how to behave.  In this case, we instruct the model to perform Source Code Summarisation.  ``Context'' gives additional contextual information that helps the model to complete the task (in the case of one-shot learning, this will include an example).  ``Input Data'' contains the input we want the model to respond to.  ``Output Indicator'' tells the model how to provide an output.

Following this prompting model, we used four prompts, described below.  A replication package for this work, including JSON examples of all prompts is linked in Appendix \ref{sec:AppendixReplication}.

\subsubsection{Zero-Shot Prompting}

For zero-shot prompting, no additional context was included in order to ensure the validity of the prompt as zero-shot.  Table \ref{table:ZeroShotPrompt} breaks the prompt down into these parts.
    \begin{table}[H]
        \centering
        \resizebox{\columnwidth}{!}
        {
            \begin{tabular}{p{.25\linewidth}p{.65\linewidth}}
                \hline
                Prompt Part & Prompt \\
                \hline
                Instruction & ``You are a source code summarisation model.'' \\
                Input data & ``When given a prompt in source code, '' + Source code \\
                Output Indicator & ``you return a one-sentence English-language summary of the code.'' \\
                \hline
            \end{tabular}
        }
        \caption{Zero-Shot Prompt Structure.}\label{table:ZeroShotPrompt}
    \end{table}

\subsubsection{One-Shot Prompting}
For the One-shot prompting with the Transformer-generated example summary in the prompt, we considered three methods of inserting the summary into the prompt.  ``Implicit'', where the same prompt is used as in the zero-shot task, but with the summary inserted as a code comment into the code.  ``Explicit'', where the model is informed that there is an example summary provided.  ``Explicit requesting improvement'', where the prompt includes a request to improve the summary provided.
\vspace{15px}

The ``Implicit'' One-shot prompt follows the same structure as the zero-shot prompt (found in Table \ref{table:ZeroShotPrompt}), but with the summary inserted as a code comment.

The structure for the ``Explicit'' prompt can be found in Table \ref{table:ExplicitOneShotPrompt}.  The major difference between ``Implicit'' and ``Explicit'' prompts is that the prompt labels the example summary given and informs the model that it exists.

    \begin{table}[H]
        \centering
        \resizebox{\columnwidth}{!}
        {
            \begin{tabular}{p{.25\linewidth}p{.65\linewidth}}
                \hline
                Prompt Part & Prompt \\
                \hline
                Instruction & ``You are a source code summarisation model. '' \\
                Context & CodeSum\textsc{Bart}-generated summary \\
                Input Data & ``When given a prompt containing an example summary and some source code,'' + Source code \\
                Output Indicator & ``you return a one-sentence English-language summary of the code.'' \\
                \hline
            \end{tabular}
        }
        \caption{Explicit One-Shot Prompt Structure.}\label{table:ExplicitOneShotPrompt}
    \end{table}

The structure for the ``Explicit requesting improvement'' prompt can be found in Table \ref{table:ExplicitReqOneShotPrompt}.

    \begin{table}[H]
        \centering
        \resizebox{\columnwidth}{!}
        {
            \begin{tabular}{p{.25\linewidth}p{.65\linewidth}}
                \hline
                Prompt Part & Prompt \\
                \hline
                Instruction & ``You are a source code summarisation model. '' \\
                Context & CodeSum\textsc{Bart}-generated summary \\
                Input Data & ``When given a prompt containing an example summary and some source code,'' + Source code \\
                Output Indicator & ``you improve that summary by returning an improved one-sentence English-language summary of the code.'' \\
                \hline
            \end{tabular}
        }
        \caption{Explicit Requesting Improvement One-Shot Prompt Structure.}\label{table:ExplicitReqOneShotPrompt}
    \end{table}

Upon finding a sizeable discrepancy between \textsc{Bleu-4} score and \textsc{Bert}Score for the results of these tests (discussed further in Section \ref{sec:Results_RQ1}), we theorised that \textsc{Bleu} may be penalising the LLM-generated responses due to their length.  We found that the human-written summaries in this dataset have both a median and mean length of 9 words, whereas LLM-generated summaries tended to be longer (a mean length of 26.64 words, and a median length of 23.67), often starting with a preamble explaining that it was a summary.

To rectify this, we added the following text to the prompts: ``Your response should be approximately 9 words long and contain only the summary.''.  The resultant summaries had a mean length of 13.12 words and a median length of 9.49 words.

\subsection{Methodology for RQ. 2}
In order to establish whether telling the LLM that the NSCS task it is performing is a form of summarisation has an effect on the quality of its outputs when measured against common NLG metrics, we repeated the methodology used in Section \ref{sec:MethodologyRQ1}, but replaced the phrases ``summary'' and ``summarisation'' with ``description'' in the prompt used.

We ran this task with two of the better-performing models from Section \ref{sec:Results_RQ1}, namely Llama 3.1 and CodeLlama, when prompted using the ```Explicit' One-shot Prompt'' - the best-performing prompt.  The augmented prompt used for this is described in Table \ref{table:ExplicitOneShotPromptNoSumm}.

    \begin{table}[H]
        \centering
        \resizebox{\columnwidth}{!}
        {
            \begin{tabular}{p{.25\linewidth}p{.65\linewidth}}
                \hline
                Prompt Part & Prompt \\
                \hline
                Instruction & ``You are a source code description model. '' \\
                Context & CodeSum\textsc{Bart}-generated summary \\
                Input Data & ``When given a prompt containing an example description and some source code,'' + Source code \\
                Output Indicator & ``you return a one-sentence English-language description of the code.'' \\
                \hline
            \end{tabular}
        }
        \caption{Explicit One-Shot Prompt Structure, with references to summarisation removed.}\label{table:ExplicitOneShotPromptNoSumm}
    \end{table}

\subsection{Methodology for RQ. 3}
To evaluate whether code-aware LLMs outperform English-language LLMs for this NSCS task, we ran this task using both general purpose English-language LLMs and Code-aware LLMs.  Our selection of LLMs can be seen in Table \ref{table:LLMs}.  We compare and contrast the evaluation results across all models in Section \ref{sec:Results_RQ1}, then further focus on the differences between code-aware and general purpose LLMs based on these results in Section \ref{sec:Results_RQ3}.

    \begin{table}[!ht]
        \centering
        \begin{tabular}{p{.49\linewidth}p{.18\linewidth}>{\raggedleft\arraybackslash}p{.16\linewidth}}
        \hline
          Large Language Model & Code Aware? & Parame-ters (Billion) \\
          \hline
          Llama 3.1 & No & 8 \\
          Llama 3.2 & No & 3 \\
          CodeLlama (based on Llama 2) & Yes & 7 \\
          Deepseek Coder 1.5 & Yes & 7 \\
          \hline
        \end{tabular}
        \caption{Large Language Models Tested.}
        \label{table:LLMs}
    \end{table}

\section{Results}
\subsection{Results Relating to RQ. 1}\label{sec:Results_RQ1}
Table \ref{table:RQ1_Results} shows the result of evaluating the four LLMs identified in Table \ref{table:LLMs} on our dataset, against \textsc{Bleu}-1\&-4, Smoothed \textsc{Bleu}-4, \textsc{Rouge}-L, and \textsc{Bert}Score.


The initial results when evaluated against \textsc{Bleu} and \textsc{Rouge} metrics are all disproportionately lower than expected for the given \textsc{Bert}Score.  The discrepancy between the two types of metric can be explained in one of two ways.  LLMs generate longer responses than the human-written summaries, meaning that metrics which include a brevity penalty penalise these as the lengths of the summaries do not match.  Also, LLMs generate highly abstractive summaries, which may be semantically similar to the summary a human would write (i.e. they share a meaning), but use different key words and phrases than the human would.  Controlling output length caused an improvement of 1.01\% \textsc{Bleu}-4 and 3.84\% \textsc{Bert}Score on a zero-shot prompt.

A small random selection of these summaries can be found in Appendix \ref{sec:AppendixExamples}.

Using the zero-shot prompting method, we observe 
similar performance between Llama 3.1 and 3.1, substantially outperforming Deepseek Coder 1.5, and being outperformed in turn by CodeLlama.  Requesting a shorter response to the prompt improved performance across all metrics and models, with CodeLlama remaining most performant.

Using the ```Implicit' One-shot'' prompt provides the same relative distribution of metrics, with Llama 3 variants performing similarly, Deepseek Coder performing poorly, and CodeLlama performing best across all evaluation Metrics.  It is noteworthy that including an example summary as a code comment causes a significant improvement in performance, with increases across all metrics - which shows that the LLMs are capable of finding usable information from the comment in the context of the provided source code method.

The ```Explicit' One-shot'' prompt provided a similar relative distribution of metrics across the board as the previous two prompts, and caused a notable improvement when compared to the ```Implicit' One-shot'' prompt.

The ``Explicit requesting improvement'' One-shot prompt we created did not cause an improvement in performance across all metrics.  While this prompt did cause an improvement in performance for Deepseek Coder 1.5 across all metrics, we observed a general decrease in performance across other models.

Prompting an LLM using our ```Explicit' One-shot'' prompting method provides the bets evaluation results for three out of four models.  We describe this method as ``Transformer-Assisted LLM-Based Source Code Summarisation''.

    \begin{table*}[!ht]
        \centering
        \resizebox{1.9\columnwidth}{!}
        {
            \begin{tabular}{rrrrrr}
              \hline
              Large Language Model & \rot{\textsc{Bleu-1}} & \rot{\textsc{Bleu-4}} & \rot{Smoothed \textsc{Bleu-4}} & \rot{\textsc{Rouge-L}} & \rot{\textsc{Bert}Score} \\
              \hline
              \multicolumn{6}{c}{Zero-shot prompt, original response length.} \\
              \hline
              Llama 3.1 & 14.96 & 1.60 & 8.14 & 26.90 & 60.94 \\
              Llama 3.2 & 9.53 & 0.71 & 4.84 & 22.24 & 58.23 \\
              CodeLlama & 13.58 & 2.07 & 7.94 & 25.84 & 57.13 \\
              Deepseek Coder 1.5 & 2.02 & 0.17 & 0.94 & 14.94 & 51.05 \\
              \hline
              \multicolumn{6}{c}{Zero-shot prompt, shortened response length.} \\
              \hline
              Llama 3.1 & 26.61 & 2.61 &17.80 & 32.80 & 64.78 \\
              Llama 3.2 & 22.71 & 1.78 & 15.05 & 28.85 & 63.52 \\
              CodeLlama & \textbf{28.60} & \textbf{4.62} & \textbf{19.19} & \textbf{36.88} & \textbf{65.79} \\
              Deepseek Coder 1.5 & 6.32 & 0.59 & 3.27 & 20.08 & 55.57 \\
              \hline
              \multicolumn{6}{c}{Implicit one-shot prompt, shortened response length.} \\
              \hline
              Llama 3.1 & 31.97 & 5.99 & 21.19 & 39.06 & 67.51 \\
              Llama 3.2 & 27.63 & 3.87 & 17.72 & 34.75 & 66.13 \\
              CodeLlama & \textbf{38.45} & \textbf{12.00} & \textbf{27.02} & \textbf{47.02} & \textbf{70.26} \\
              Deepseek Coder 1.5 & 8.05 & 2.02 & 4.71 & 24.08 & 57.56 \\
              \hline
              \multicolumn{6}{c}{Explicit one-shot prompt, shortened response length.} \\
              \hline
              Llama 3.1 & 34.57 & 7.42 & 23.33 & 41.61 & 68.69 \\
              Llama 3.2 & 28.08 & 4.22 & 18.33 & 34.83 & 66.26 \\
              CodeLlama & \textbf{39.96} & \textbf{12.42} & \textbf{28.57} & \textbf{48.04} & \textbf{70.79} \\
              Deepseek Coder 1.5 & 18.89 & 5.77 & 12.13 & 32.40 & 62.46 \\
              \hline
              \multicolumn{6}{c}{Explicit Requesting Improvement one-shot prompt, shortened response length.} \\
              \hline
              Llama 3.1 & 32.86 & 5.45 & 21.63 & 39.43 & 67.79 \\
              Llama 3.2 & 28.65 & 4.01 & 18.75 & 35.10 & 66.36 \\
              CodeLlama & \textbf{38.95} & \textbf{11.50} & \textbf{27.50} & \textbf{47.13} & \textbf{70.37} \\
              Deepseek Coder 1.5 & 27.37 & 8.47 & 18.18 & 38.69 & 66.02 \\
              \hline
            \end{tabular}
        }
        \caption{Evaluation of LLM Outputs After Prompting.}\label{table:RQ1_Results}
    \end{table*}

\subsection{Results Relating to RQ. 2}\label{sec:Results_RQ2}
Table \ref{table:RQ2_Results} shows that removing references to summarisation has a slight positive impact across all metrics for the CodeLlama model, and across five of the six metrics for Llama 3.1.  These results show that the choice of prompt wording has some impact on the quality of model predictions, but the impact of this is limited in this case, when compared to the previous best results, found in Table \ref{table:RQ1_Results}.

    \begin{table*}[!ht]
        \centering
        \begin{tabular}{rrrrrrr}
          \hline
          Large Language Model & \rot{\textsc{Bleu-1}} & \rot{\textsc{Bleu-4}} & \rot{Smoothed \textsc{Bleu-4}} & \rot{\textsc{Rouge-L}} & \rot{\textsc{Bert}Score} \\
          \hline
          Llama 3.1 & 33.71 & 7.85 & 22.97 & 40.97 & 68.60 \\
          CodeLlama & 40.61 & 12.80 & 29.08 & 48.63 & 71.13 \\
          \hline
        \end{tabular}
        \caption{Explicit One-Shot Prompt, shortened responses, with references to summarisation removed.}
        \label{table:RQ2_Results}
    \end{table*}

\subsection {Results relating to RQ. 3}\label{sec:Results_RQ3}
Across all prompts and all metrics, shown in Table \ref{table:RQ1_Results}, CodeLlama outperformed all General-purpose LLMs.  The increased performance by CodeLlama is especially notable when compared to Llama 3.1, which has a greater number of parameters than CodeLlama.

However, The General-purpose LLMs all outperform Deepseek Coder 1.5, which has the same number of parameters as CodeLlama.  The difference between the two could be attributed to the different model architectures requiring different prompting methods (as seen in the results of the ``Explicit Requesting Improvement'' prompt, where other models showed a drop off in performance by a prompt change which improved the performance of Deepseek Coder), or could be attributed to the difference between the pretraining dataset and methods used by Deepseek Coder and that of CodeLlama.

CodeLlama's performance shows that code-aware LLMs have the potential to significantly outperform English-language LLMs for NSCS tasks, however, Deepseek Coder's performance shows that this improvement is not universal.  Further work is needed to quantify how much improvement code-aware LLMs can present for NSCS tasks on a larger scale.

\section{Implications for Secure Software Development}
In this paper, we work towards solving an issue in secure software development: a lack of source code summaries, or documentation in a wider context, makes code more difficult to comprehend, which in turn could negatively affect maintainers' ability to identify bugs and vulnerabilities in source code.  \citet{xia-etal-2017-measuring} found that developers ``cannot understand the source code if there are insufficient comments[...]''.  A lack of documentation presents a major risk in secure development, which this paper seeks to address by generating source code summaries using LLMs.

The use of LLMs as part of a secure software development stack is a complex and debated task, as shown in work by \citet{yao-etal-2024-survey}.  In order to minimise the ``black box'' effect of using LLMs (i.e.: being unable to see how an LLM works and handles data), we selected only LLMs which are open source and available via HuggingFace\footnote{https://huggingface.co/}.  It should be noted, however, that in order to use Deepseek Coder 1.5 via HuggingFace's API, it is required to set the \texttt{ trust\_remote\_code$=$True} flag, which allows remote code execution.  While the code which is being executed can be viewed and reviewed for trustworthiness, allowing remote code execution as part of a software development stack has the potential to pose a risk to secure development.  not only does using LLMs which execute remote code present potential safety implications by running code which has not been checked for vulnerabilities and could - theoretically - be used maliciously, but using LLMs to generate source code summaries should be done with developer oversight, as these models have the potential to ``hallucinate'' and give incorrect summaries.  

\section{Conclusion}
We find that LLMs are capable of generating summaries of source code methods in a manner which maintains a high degree of semantic similarity to human-written summaries, as reflected in a high \textsc{Bert}Score value.

We present ``Transformer-Assisted LLM-Based Source Code Summarisation'' - a method of one-shot prompting a Large Language Model for NSCS tasks, which can be used with a number of LLMs.  CodeLlama performs exceptionally well when prompted using this method, with a \textsc{Bleu}-1 score of 39.96\% and a \textsc{Bert}Score of 70.79\% on our dataset.

We find that prompting a Large Language Model with the output of a smaller
transformer model improves the summary quality when evaluated against NLG metrics on an NSCS task, showing across-the-board improvements with four different LLMs across 5 metrics.

Referencing summarisation as a part of the task has a small impact on the quality of the summary; replacing key words such as ``summarisation'' and ``summary'' with similar terms, such as ``description'' appears to cause a small improvement to the model predictions (LLM-generated summaries) when measured with a selection of NLG metrics.

Code-aware LLMs show promise when compared to English-language LLMs on an NSCS task, however, model selection and pretraining is pivotal to seeing this improvement and not all code-aware LLMs are capable of this improvement, as discussed in Section \ref{sec:Limitations}. 

\section*{Limitations}\label{sec:Limitations}
We show one area where improvements can be made to LLM prompting for NSCS tasks, however, we have used a small set of LLMs to show this.  A larger set of LLMs could be tested to validate these findings across a wider array of LLM architectures.  Most notably, all LLMs used for this study are small models with $< 10$ Billion parameters, such that they can be run on a powerful workstation PC.  All models were prompted using a single \texttt{Nvidia A5000 24GB} GPU, so future work is needed to investigate the scaling of these findings on larger models.  In addition, future work is needed to establish developer preference for these types of summary.

The two code-aware LLMs used in this study provided vastly different results despite their similar size, and future work is needed to investigate whether a general trend can be found either in support of or against the use of code-aware LLMs for NSCS tasks, rather than English-language LLMs - however this initial evidence suggests that code-aware LLMs show promise for NSCS use in future.  The selection of LLMs for this paper is a limiting factor, and future work analysing the outputs of recent frontier models and other code-aware LLMs when prompted using our approach could provide further improvements.

The use of the Funcom dataset presents one potential limitation, in terms of generalisability.  While the Funcom dataset is commonly used for evaluating NSCS models, the methods contained within the dataset were collected in 2010, meaning that there is potential that this dataset may not be reflective on current trends in software development.  As well as the dataset used presenting one potential threat to the generalisability of this work, the metrics used present another.  As this has only been evaluated against a suite of NLG metrics, further work is needed to evaluate these methods using human evaluators.

\section*{Ethics Statement}
There are two main ethical implications of this work.  Firstly, the use of the Funcom dataset \citep{leclair-mcmillan-2019-recommendations}, which is derived from \citet{lopes-etal-2010-uci}'s UCI dataset of \num{18000} Java projects from 2010.  The dataset is no longer available in its original format, which was hosted at \href{http://leclair.tech/data/funcom/}{leclair.tech/data/funcom}, and is only available through either having downloaded a copy of the dataset when it was available, or through other online mirrors of the dataset.  The licensing of this data is unclear, but is assumed to follow the license terms of the UCI Source Code dataset from which it is built \citep{lopes-etal-2010-uci}.

Secondly, the environmental impact of prompting and using LLMs.  We attempted to minimise this impact by selecting LLMs with $<10$ Billion parameters, and prompting them using a single \texttt{Nvidia A5000 24GB} GPU, however all work using LLMs has some environmental impact.

\section*{Acknowledgements}
This research was made possible through the use of ``Hex'' - A high-performance compute cluster created by the UCREL NLP group at Lancaster University.  \citep{vidler-rayson-2024-ucrel}

\bibliography{anthology,custom}
\bibliographystyle{acl_natbib}

\appendix

\section{Update to JavaDatasetCleaner}\label{sec:AppendixJava}
\begin{lstlisting}[language=Java, label={lst:UpdatedJavaDatasetPreprocessor.java}, keywordstyle=\color{blue}, stringstyle=\color{red}, breaklines=true, backgroundcolor=\color{gray!10}, caption={Updated method added to JavaDatasetPreprocessor.java},captionpos=b,basicstyle=\linespread{0.8}\ttfamily\footnotesize]
/**
* Re-tokenises summaries.
**/
public void retokeniseSummaries()
{
  for (int i = 0; i < summaries.size(); i++)
  {
    // Split the string on camel case.
    String[] camelSplitData = summaries.get(i).replaceAl1("_", " ").split("(?<!(^|[A-Z]))(?=[A-Z])|(?<!^)(?=[A-Z][a-z])");
    StringBuilder tmp = new StringBuilder();
    for (String c : camelsplitData)
      tmp.append(" ").append(c);
    summaries.set(i, tmp.toString().trim());
  }
  // Replace "." with " . " and "," with " , " to tokenise the words around the punctuation.
  summaries.replaceAll (s1 -> s1.replaceA11("\\.", " . "));
  summaries. replaceAll(s1 -> s1.replaceA11(",", " , "));
  // Replace multiple spaces with a single space.
  summaries.replaceAll(s1 -> s1.trim().replaceA11("\\s+", " "));
  // Pad out apostrophes being used as single quotes.
  summaries.replaceAll(s1 -> s1.trim().replaceA11(" \\'", " ' "));
  summaries.replaceAll(s1 -> s1.trim().replaceAll("\\' ", " ' "));
}
\end{lstlisting}

\section{Replication Package}\label{sec:AppendixReplication}
A replication package for this work can be found at: 
\href{https://github.com/phillijm/TransformerAssistedLLMCodeSum}{github.com/phillijm/\\TransformerAssistedLLMCodeSum}.
\onecolumn
\section{Example Outputs}\label{sec:AppendixExamples}
These examples use CodeLlama, with a shortened response length.
\begin{lstlisting} [breaklines=true]
Example 1:
    Human-written summary: Returns a string describing this ring.
    Zero-shot summary: The code converts a linked list to a string by iterating through the list and concatenating the values of each node with a space in between.
    Implicit One-shot summary: Returns a string representation of the ring by iterating through the elements and concatenating them with spaces.
    Explicit One-shot summary: Returns a string representation of the ring by iterating over its elements and concatenating them with spaces.
    Explicit Requesting Improvement One-shot summary: Returns a string representation of the ring by iterating over its elements and concatenating them with spaces.

Example 2:
    Human-written summary: Writes the modified field tree.
    Zero-shot summary: Writes data tree to output stream.
    Implicit One-shot summary: Writes the tree class tree to the specified writer.
    Explicit One-shot summary: Writes the tree class tree to the writer out. 
    Explicit Requesting Improvement One-shot summary: Writes the tree class tree to the output stream.

Example 3:
    Human-written summary: Setter method for javascript onlick
    Zero-shot summary: Sets the onclick property to a new value.
    Implicit One-shot summary: Sets the onclick attribute to a new value.
    Explicit One-shot summary: Sets the value of the onclick attribute.
    Explicit Requesting Improvement One-shot summary: Sets the value of the onclick attribute.

Example 4:
    Human-written summary: Tests the default values set by the builder.
    Zero-shot summary: Tests default values of CharacterStreamDescriptor.
    Implicit One-shot summary: The code tests the default values of a character stream descriptor.
    Explicit One-shot summary: The code tests the default values of a character stream descriptor.
    Explicit Requesting Improvement One-shot summary: Tests the default values of a character stream descriptor.
\end{lstlisting}

\end{document}